\documentclass[11pt]{article}
\usepackage[margin=0.25in]{geometry}
\usepackage{authblk}

\usepackage{amsmath, amsfonts, amsthm}
\usepackage{threeparttable, graphicx}
\usepackage{multirow}
\usepackage{hyperref}
\usepackage[table]{xcolor}
\usepackage{booktabs,tabulary}
\usepackage{siunitx}

\usepackage[english]{babel}

\begin{document}

\title{Estimating the Distribution of Ratio of Paired Event Times in Phase II Oncology Trials} 
\date{}
\author[1,2]{Li Chen \thanks{Corresponding Author. Address: 800 Rose St., CC425, Lexington, KY 40536, United States. Email: lichenuky@uky.edu}}

\author[3]{Mark Burkard}

\author[1,2]{Jianrong Wu}

\author[1,4]{Jill M. Kolesar}

\author[1,2]{Chi Wang}

\affil[1]{Biostatistics and Bioinformatics Shared Resource Facility, Markey Cancer Center, University of Kentucky, Kentucky, Unites States}
\affil[2]{Department of Internal Medicine, University of Kentucky, Kentucky, Unites States}
\affil[3]{Department of Medicine, University of Wisconsin—Madison, Wisconsin, Unites States}
\affil[4]{Department of Pharmacy Practice \& Science, University of Kentucky, Kentucky, Unites States}

\begin{titlepage}
\maketitle
\end{titlepage}


\abstract{With the rapid development of new anti-cancer agents which are cytostatic, new endpoints are needed to better measure treatment efficacy in phase II trials. For this purpose, Von Hoff (1998) proposed the growth modulation index (GMI), i.e. the ratio between times to progression or progression-free survival times in two successive treatment lines. An essential task in studies using GMI as an endpoint is to estimate the distribution of GMI. Traditional methods for survival data have been used for estimating the GMI distribution because censoring is common for GMI data. However, we point out that the independent censoring assumption required by  traditional survival methods is always violated for GMI, which may lead to severely biased results. In this paper, we construct  nonparametric estimators for the distribution of GMI, accounting for the dependent censoring of GMI. We prove that the proposed estimators are consistent and converge weakly to zero-mean Gaussian processes upon proper normalization. Extensive simulation studies show that our estimators perform well in practical situations and outperform traditional methods. A phase II clinical trial using GMI as the primary endpoint is  provided for illustration.
}


\maketitle


\section{Introduction}\label{intro}
With tremendous advances in cancer biology, targeted therapies have emerged as a new generation of cancer treatment that uses drugs to target specific molecules involved in the growth and spread of cancers \cite{mellman2011cancer, a2015targeted, talwar2017overall}. This evolution in drug development has demanded innovation in the design and endpoint selection of phase II trials to more effectively evaluate the efficacy of a new targeted therapy. Unlike conventional cytotoxic  agents, targeted therapies are often cytostatic, acting to block tumor growth but do not lead to an immediate tumor shrinkage \cite{serkova2016metabolic, anttila2019contrasting}. Thus, objective tumor response, which is the primary endpoint for phase II trials on cytotoxic agents, is inappropriate for targeted therapy trials \cite{ritchie2018defining}. Because the cytostatic agents are expected to delay disease progression,  trial designs based on the  time to progression (TTP) and progression-free survival (PFS) are preferable \cite{ritchie2018defining}. TTP is defined as the time from the start of treatment to the event of tumor progression. PFS is closely related to TTP except that it also considers death as the event of interest. 

Von Hoff (1998)\cite{von1998there} proposed the growth modulation index (GMI) as a primary endpoint to evaluate treatment efficacy for cytostatic agents. The GMI is defined as the intra-patient TTP (or PFS) ratio, i.e. the ratio of the TTP (or PFS) on the new treatment versus that on the most recent treatment on which the patient had experienced progression. By using each patient as his/her own control, the GMI directly measures the delay in disease progression. Because of this appealing feature, it has been increasingly used as the primary endpoint in phase II trials for targeted therapies \cite{von2010pilot, radovich2016clinical, rodon2019genomic, sicklick2019molecular}.


An essential problem in studies using GMI as an endpoint is to estimate the probability of GMI greater than a given value, which is referred to as the survival function of GMI. For example, in a phase II study on genomical profiling guided treatments conducted by Von Hoff \cite{von2010pilot}, the probability of GMI $>$ 1.3 is of particular interest because patients with GMI $>$ 1.3 were considered to have a meaningful improvement in PFS. Hypothesis testing for treatment efficacy was further performed to test whether that probability was greater than a pre-specified proportion.

To estimate the survival function of GMI, it is important to notice that GMI  is often censored in practice. Let $T_0$ and $T_1$ be TTPs (or PFSs) on the new and prior treatments, respectively. The $T_0$ is observed since all patients enrolled in the trial had progressed on prior treatment. The $T_1$ may be right censored. Let $C_1$ be the censoring time.  Thus, the GMI, i.e. $T_1 / T_0$, is  right censored by $C_1/T_0$. Note that $T_1 / T_0$ and $C_1/T_0$ are dependent even when $T_1$ and $C_1$ are independent because they both involve $T_0$.  Therefore, GMI is always dependently censored. To our knowledge, this dependent censoring issue has not been recognized in the literature. Current methods for estimating the survival function of GMI are based on traditional methods for right-censored survival data, such as the Kaplan-Meier (KM) method and parametric survival models\cite{texier2018evaluation, kovalchik2011statistical}. However, those methods require the independent censoring assumption, which does not hold for GMI. Thus, they may lead to biased results for estimating the survival funciton of GMI. No methods have been developed to account for the dependent censoring of GMI. 

In this paper, we develop consistent nonparametric estimators for the survival function of GMI accounting for the dependent censoring of GMI. We consider both  the situation that $T_1$ and $C_1$ are independent and the situation that $T_1$ and $C_1$ are independent given covariates, which are two commonly used independent censoring assumptions for event times. We notice that GMI and the corresponding censoring variable ($C_1/T_0$) are independent given $T_0$ (and covariates). Therefore, we propose to first use the kernel conditional KM method to consistently estimate the conditional survival function of GMI, and then take the average of the estimated conditional survival functions across $T_0$ (and covariates) values for all individuals. In Section 2, we propose the estimators and prove that they are consistent and weakly converge to zero-mean Gaussian processes upon proper normalization. In Section 3, we conduct extensive simulation studies to evaluate the performance of the proposed estimators and compare to traditional methods. In Section 4, analysis of data from a phase II clinical trial is presented for illustration. Finally, we conclude this paper with a discussion and possible future work in Section 5.

\section{Methods}\label{sec2}
Let $Y_1 = \min(T_1, C_1)$ and $\Delta_1 = I(T_1 \leq C_1)$. The data consist of $n$ independent replicates $\{T_{0i}, Y_{1i}, \Delta_{1i}\}$, $i = 1\ldots,n$. Our goal is to develop consistent nonparametric estimators for the survival function of GMI, i.e. $S_{{\rm G}}(r) = {\rm Pr}(T_1/T_0 > r)$. 


We first consider the situation that $T_1$ and $C_1$ are independent. Under this situation, we notice that given $T_0$, $T_1/T_0$ and $C_1/T_0$ are independent. Thus, the conditional survival function of GMI given $T_0=t_0$, i.e. $S_{{\rm G}|T_0}(r; t_0) = {\rm Pr}(T_1/T_0 > r | T_0=t_0)$, can be consistently estimated by the kernel conditional KM method. Further, the survival and  conditional survival functions of GMI are connected by the formula of $S_{{\rm G}}(r) = E_{T_0}\{S_{{\rm G}|T_0}(r; T_0)\}$  Therefore, to consistently estimate $S_{{\rm G}}(r)$, we first estimate the conditional survival function of GMI given $T_0$, and then take the average of the estimated conditional survival function over $T_0$ values for all individuals. Specifically, we propose to estimate $S_{{\rm G}|T_0}(r; t_0)$ by the following kernel conditional KM type estimator
\begin{equation}\label{eq:ckm}
\hat{S}_{{\rm G}|T_0}(r; t_0) = \Pi_{0<s\leq r} \left[1-\frac{\sum_{j=1}^n K\left\{\left(\log(T_{0j}) - \log(t_{0})\right)/{a_n}\right\} I(Y_{1j}/T_{0j}=s)\Delta_{1j}}{\sum_{j=1}^n K\left\{\left(\log(T_{0j}) - \log(t_0)\right)/{a_n}\right\} I(Y_{1j}/T_{0j}\geq s)}\right],
\end{equation}
where $K(\cdot)$ is a kernel function and $a_n$ is the bandwidth. We then estimate $S_{{\rm G}}(r)$ by 
\begin{equation}
\hat{S}_{{\rm G}}(r) = \frac{1}{n}\sum_{i=1}^n \hat{S}_{{\rm G}|T_0}(r; T_{0i}).
\end{equation}
Note that the kernel function $K(\cdot)$ appears in the denominator of formula (\ref{eq:ckm}). So extremely small kernel weight values should be avoided for computational stability. Because the distribution of $T_{0j}$ is often right skewed, there may be very large values of $T_{0j}$ that could result in extremely small kernel weight values. To solve this problem, we use $\log(T_{0j})$ instead of $T_{0j}$ in the kernel calculation. Further, we choose to use the following modified Silverman kernel \cite{silverman1986density, yang2017estimation} rather than  the standard Gaussian kernel because it is flatter and thus less likely to produce extremely small kernel weight
\begin{eqnarray*}
K(u)=\frac{ |\frac{1}{2}e^{\frac{-|u|}{\sqrt{2}}}\sin(\frac{|u|}{\sqrt{2}} + \frac{\pi}{4})| }
{ \int_{-\infty}^{\infty}
|\frac{1}{2}e^{\frac{-|u|}{\sqrt{2}}}\sin(\frac{|u|}{\sqrt{2}} + \frac{\pi}{4})|du}.
\end{eqnarray*}

We show in Appendix that under the conditions that $K(\cdot)$ is a symmetric smooth probability density function and $na_n\rightarrow \infty$ and $na_n^4\rightarrow 0$, $\hat{S}_{{\rm G}}(r)$ is consistent and $\sqrt{n}(\hat{S}_{{\rm G}}(r)-S_{{\rm G}}(r))$ converges weakly to a zero-mean Gaussian process and is asymptotically equivalent to the process $n^{-1/2}\sum_{i=1}^n \xi_i(r)$, where
\begin{equation}\label{var}
\xi_i(r) =  S_{{\rm G}| T_0}(r; T_{0i})\left\{1-\frac{\Delta_{1i}I(Y_{1i}/T_{0i} \leq r)}{H(Y_{1i}/T_{0i}; T_{0i})}  - \int_0^r\frac{I(Y_{1i}/T_{0i} \leq s)d\log S_{{\rm G}| T_0}(s; T_{0i})}{H(s; T_{0i})}\right\} - S_{\rm G}(r) ,
\end{equation}
with $H(s; t_0) = {\rm Pr}(Y_{1}/T_{0}\geq s|T_0=t_0)$.

The variance of the estimator $\hat{S}_{{\rm G}}(r)$
can be estimated by ${n}^{-2}\sum_{i=1}^{n}\hat\xi_i^2(r)$, where $\hat\xi_i(r)$ is obtained by replacing 
$S_{{\rm G}|T_0}(\cdot; \cdot)$, $S_{{\rm G}}(\cdot)$ and $H(\cdot; \cdot)$ by their estimators in equation (\ref{var}). However, this formula-based variance estimation method may be unstable for small sample sizes. Therefore, we recommend using the bootstrap method to calculate the variance estimator. The log-log transformed 95\% confidence intervals for $S_{{\rm G}}(r)$ can then be calculated.

Next, we extend the above estimator to the situation that $T_1$ and $C_1$ are allowed to be dependent but are required to be independent given $Z$ and $V$, where  $Z$ is a set of continuous covariates and $V$  is a set of categorical covariates. Under this assumption, $T_1/T_0$ and $C_1/T_0$ are independent given $T_0$, $Z$ and $V$. Analogous to the first situation, the survival function of GMI is estimated by

\begin{equation*}
\tilde{S}_{{\rm G}}(r) = \frac{1}{n}\sum_{i=1}^n \tilde{S}_{{\rm G}|T_0, Z, V}(r; T_{0i}, Z_i, V_i),
\end{equation*}
where
\begin{equation*}
\tilde{S}_{{\rm G}|T_0,Z, V}(r; t_0,z,v) = \Pi_{0<s\leq r} \left[1-\frac{\sum_{j: V_j=v} K\left\{\left\|\left(X_j - x\right)/{a_n}\right\|\right\} I(Y_{1j}/T_{0j}=s)\Delta_{1j}}{\sum_{j: V_j=v} K\left\{\left\|\left(X_j - x\right)/{a_n}\right\|\right\} I(Y_{1j}/T_{0j}\geq s)}\right],
\end{equation*}
 $Z_j$ and $V_j$ are $Z$ and $V$ values for subject $j$, respectively, $X_j = (\log(T_{0j}), Z_j)$, $x = (\log(t_0), z)$, $\|.\|$ is the Euclidean norm, $K(\cdot)$ is a kernel function, and $a_n$ is the bandwidth. As shown in Appendix, under the conditions that 1)
there exists some integer $l$ such that $\int u^{j} K(u)du=0$ ($j=1, \ldots, l-1$),
$\int u^{l} K(u)du\neq 0$ and $2l-q-1>0$, where $q$ is the dimension of $Z$; and 2)
$na_n^{q+1}\rightarrow \infty$ and $na_n^{2l}\rightarrow 0$ as $n\rightarrow\infty$, $\tilde{S}_{{\rm G}}(r)$ is consistent and $\sqrt{n}\{\tilde{S}_{{\rm G}}(r) - S_{{\rm G}}(r)\}$ is weakly convergent to a zero-mean Gaussian process. When $q \leq 2$, a symmetric smooth probability density function can be used as the kernel function. When $q > 2$, due to the difficulty in choosing an appropriate kernel function and the curse of dimensionality, we suggest obtaining a linear combination of $Z$ under a proportional hazards model for censoring time $C_1$, and then applying the proposed estimator $\tilde{S}_{{\rm G}}(r)$  to the linear combination of $Z$ instead of $Z$.

\section{Simulation Studies}\label{sec3}

We assessed the performance of the proposed estimator, $\hat{S}_{{\rm G}}(\cdot)$, and compare to the KM estimator and parametric estimators based on  lognormal and loglogistic distributions. For $\hat{S}_{{\rm G}}(\cdot)$, the bandwidth of $\hat{\sigma}_0 n^{-2/5}$ was used, where $\hat{\sigma}_0$ was the sample standard deviation of $\log T_0$. For the KM estimator, the survfit function in R was used for the calculation. For parametric estimators, the survreg function in R was used to obtain the estimators. We generated paired event times $T_0$ and $T_1$ from a Weibull frailty model, i.e. $T_0 \sim {\rm Weibull}(e^\mu\theta, \sigma^{-1})$ and $T_1 \sim {\rm Weibull}(e^\mu\theta R, \sigma^{-1})$ given the frailty term $\theta$, where $\theta$ follows a Gamma distribution with both shape and rate parameters equal to $\alpha$. Marginally, $T_0 \sim {\rm Weibull}(e^\mu, \sigma^{-1})$ and $T_1 \sim {\rm Weibull}(e^\mu R, \sigma^{-1})$, where $R$ is the ratio of medians of $T_1$ vs. $T_0$. We considered $R=$ 1 or 1.3. The $R=1$ refers to the situation that there is no improvement in the new therapy compared to the prior therapy, and $R=1.3$ refers to a situation that the new therapy is more effective than the prior therapy.  The $\mu$ was set to 3.0, which was obtained from fitting a Weibull distribution to the data of $T_0$ in the phase II clincial trial dataset in Section 4. The  $\sigma$ was set to 0.3 or 0.5, where the value of 0.5 was also obtained from the phase II trial data in Section 4 and the value of 0.3 was additionally considered for the situation of less variation.  The $\alpha$ was set such that the correlation between $T_0$ and $T_1$ equal to 0.3 or 0.5. The censoring time $C_1\sim {\rm Uinf}(0.85\tau, \tau)$, where $\tau$ was set such that the censoring rate was 20\% or 30\%. We considered sample size $n = 50, 70, 90$ and 2000 simulation replicates.

	\begin{threeparttable}[h!]
		\caption{\small Simulation results for estimators of the survival function of GMI, i.e. $S_{{\rm G}}(r)$, where $T_1$ and $T_0$ were generated from a Weibull frailty model with a correlation of 0.5 and ratio of medians equal to 1. \label{table:simu1}}%
		{\scriptsize \begin{tabular*}{\linewidth}{@{\extracolsep\fill}llllllllllllllllllllllllllllll@{\extracolsep\fill}}
\cline{1-20}
&&&& \multicolumn{4}{c}{Proposed} & \multicolumn{4}{c}{KM} & \multicolumn{4}{c}{lognormal} & \multicolumn{4}{c}{loglogistic} \\
$\sigma$ &  r  & C & n & Bias & SE & SEE & CP & Bias & SE & SEE & CP & Bias & SE & SEE & CP & Bias & SE & SEE & CP\\
 \cline{1-20} .3 & 1.3 &20\% &50 & .016 & .075 & .074 & .94 & .042 & .074 & .071 & .91 & .062 & .064 & .061 & .83 & .042 & .065 & .064 & .9\\
 &  & &70 & .013 & .062 & .062 & .95 & .041 & .06 & .06 & .91 & .062 & .053 & .052 & .78 & .042 & .054 & .054 & .89\\
 &  & &90 & .015 & .056 & .055 & .95 & .043 & .054 & .053 & .88 & .064 & .047 & .046 & .73 & .044 & .048 & .047 & .86\\
 &  &30\% &50 & .03 & .082 & .081 & .94 & .067 & .079 & .076 & .88 & .086 & .068 & .066 & .76 & .066 & .07 & .069 & .85\\
 &  & &70 & .026 & .068 & .069 & .94 & .065 & .065 & .065 & .84 & .086 & .056 & .055 & .67 & .067 & .058 & .058 & .81\\
 &  & &90 & .026 & .063 & .061 & .93 & .067 & .058 & .057 & .79 & .088 & .051 & .049 & .58 & .068 & .053 & .051 & .75\\
\cline{3-20} & 1.5 &20\% &50 & .018 & .069 & .068 & .94 & .047 & .069 & .067 & .91 & .064 & .061 & .059 & .81 & .042 & .059 & .058 & .9\\
 &  & &70 & .014 & .058 & .058 & .95 & .044 & .058 & .057 & .88 & .064 & .052 & .05 & .75 & .042 & .049 & .049 & .87\\
 &  & &90 & .015 & .052 & .051 & .94 & .047 & .051 & .051 & .86 & .066 & .046 & .044 & .68 & .043 & .044 & .044 & .85\\
 &  &30\% &50 & .032 & .079 & .077 & .94 & .073 & .076 & .074 & .84 & .088 & .067 & .064 & .73 & .066 & .066 & .065 & .84\\
 &  & &70 & .028 & .065 & .065 & .94 & .07 & .063 & .063 & .8 & .089 & .056 & .055 & .64 & .065 & .055 & .055 & .79\\
 &  & &90 & .028 & .06 & .058 & .92 & .072 & .057 & .056 & .74 & .09 & .051 & .049 & .52 & .066 & .05 & .048 & .72\\
\cline{3-20} & 1.7 &20\% &50 & .017 & .063 & .062 & .95 & .046 & .064 & .063 & .89 & .058 & .057 & .054 & .81 & .038 & .052 & .051 & .9\\
 &  & &70 & .013 & .054 & .052 & .94 & .044 & .055 & .053 & .86 & .058 & .049 & .046 & .75 & .038 & .044 & .043 & .86\\
 &  & &90 & .014 & .048 & .047 & .94 & .046 & .048 & .047 & .83 & .06 & .044 & .042 & .68 & .039 & .039 & .038 & .84\\
 &  &30\% &50 & .032 & .074 & .072 & .93 & .073 & .072 & .071 & .82 & .082 & .064 & .061 & .72 & .061 & .06 & .058 & .84\\
 &  & &70 & .027 & .063 & .061 & .92 & .07 & .062 & .06 & .77 & .082 & .054 & .052 & .63 & .06 & .05 & .049 & .78\\
 &  & &90 & .025 & .058 & .054 & .92 & .071 & .055 & .054 & .72 & .083 & .049 & .046 & .53 & .06 & .045 & .044 & .72\\
 \cline{1-20} .5 & 1.3 &20\% &50 & .015 & .078 & .076 & .94 & .035 & .076 & .073 & .93 & .053 & .063 & .061 & .87 & .038 & .066 & .066 & .93\\
 &  & &70 & .014 & .064 & .064 & .96 & .035 & .062 & .062 & .92 & .054 & .053 & .052 & .82 & .04 & .055 & .055 & .9\\
 &  & &90 & .014 & .058 & .057 & .95 & .037 & .055 & .055 & .91 & .055 & .046 & .046 & .79 & .041 & .049 & .049 & .87\\
 &  &30\% &50 & .028 & .084 & .082 & .94 & .057 & .08 & .077 & .89 & .075 & .067 & .065 & .8 & .061 & .071 & .069 & .87\\
 &  & &70 & .027 & .068 & .07 & .95 & .057 & .065 & .065 & .88 & .077 & .055 & .055 & .72 & .063 & .058 & .058 & .83\\
 &  & &90 & .027 & .062 & .062 & .94 & .059 & .058 & .057 & .85 & .078 & .048 & .048 & .66 & .064 & .051 & .051 & .79\\
\cline{3-20} & 1.5 &20\% &50 & .016 & .075 & .074 & .96 & .039 & .073 & .072 & .93 & .06 & .063 & .061 & .84 & .041 & .065 & .064 & .92\\
 &  & &70 & .014 & .061 & .063 & .96 & .039 & .06 & .061 & .91 & .061 & .053 & .052 & .78 & .042 & .055 & .054 & .89\\
 &  & &90 & .016 & .057 & .056 & .94 & .041 & .055 & .054 & .89 & .062 & .047 & .046 & .74 & .043 & .048 & .048 & .85\\
 &  &30\% &50 & .03 & .085 & .081 & .93 & .063 & .08 & .076 & .87 & .083 & .068 & .066 & .77 & .065 & .071 & .069 & .86\\
 &  & &70 & .028 & .067 & .069 & .94 & .063 & .065 & .065 & .85 & .085 & .057 & .055 & .67 & .066 & .059 & .058 & .81\\
 &  & &90 & .029 & .062 & .061 & .93 & .065 & .058 & .057 & .81 & .086 & .049 & .049 & .6 & .067 & .051 & .051 & .77\\
\cline{3-20} & 1.7 &20\% &50 & .016 & .073 & .072 & .95 & .042 & .072 & .07 & .92 & .063 & .063 & .061 & .82 & .041 & .063 & .062 & .91\\
 &  & &70 & .016 & .061 & .061 & .95 & .042 & .06 & .059 & .9 & .064 & .053 & .051 & .76 & .042 & .053 & .052 & .88\\
 &  & &90 & .016 & .054 & .054 & .95 & .044 & .053 & .052 & .88 & .066 & .047 & .046 & .71 & .044 & .047 & .046 & .84\\
 &  &30\% &50 & .032 & .083 & .08 & .93 & .068 & .079 & .076 & .86 & .087 & .068 & .065 & .74 & .066 & .069 & .067 & .85\\
 &  & &70 & .03 & .068 & .068 & .94 & .067 & .065 & .064 & .83 & .088 & .057 & .055 & .64 & .067 & .058 & .057 & .79\\
 &  & &90 & .031 & .062 & .061 & .92 & .07 & .058 & .057 & .78 & .09 & .05 & .049 & .56 & .068 & .05 & .05 & .75\\
 \cline{1-20}
		\end{tabular*}}
		\begin{tablenotes}
			\item {\small Note: C, censoring rate; Bias, the sampling bias; SE, the sampling standard error; SEE, the sampling mean of the standard error estimator;
CP, the coverage probability of the 95\% confidence interval. Each entry was based on 2000 repliates. SEEs for the proposed estimator and parametric estimators were calculated based on 5000 bootstrap samples.}
		\end{tablenotes}
	\end{threeparttable}
\\\\

Tables \ref{table:simu1} and \ref{table:simu2} display the results for the scenarios of median ratio $R=1$ and 1.3 , respectively, with the correlation between $T_0$ and $T_1$ equal to 0.5. For the scenario of $R=1$ (Table \ref{table:simu1}), our method performs well in all situations. Our estimator has small biases and its standard error estimator reflects the true variation. The 95\% confidence interval has coverage probabilities close to the nominal value of 0.95. In contrast, the biases of KM and parametric methods are large. Specifically, the biases of KM, loglogistic and lognormal methods are 2.1 to 3.4 times, 1.9 to 3.2 times, and 2.5 to 4.7 times as large as that of our method, respectively. Coverage probabilities of 95\% confidence intervals from KM and parametric methods are poor as the majority of them are $\leq 0.9$. The coverage probabilities become worse when the sample size and censoring rate increase and $\sigma$ decreases. Particularly, for the case of sample size of 90, censoring rate of 30\% and $\sigma$ of 0.3, the coverage probabilities at threshold value of 1.3 are as low as 0.79, 0.75 and 0.58 for the KM, loglogistic and lognormal  methods, respectively. For the scenario of $R=1.3$ (Table \ref{table:simu2}), our method also performs well with small biases and appropriate coverage probabilities. The biases of all other methods are still large compared to our method. The biases of KM, loglogistic and lognormal methods are 2.0 to 3.0 times, 2.2 to 3.5 times, and 2.5 to 4.6 times as large as that of our method, respectively. As for the coverage probability, the KM method has coverage probabilites $\leq 0.9$ for more than half of the cases at the threshold of 1.7, for almost all the cases with a censoring rate of 30\% at threshold of 1.5, and for all cases with a censoring rate of 30\% and sample size of 90 at threshold of 1.3. The parametric methods have poor coverage probabilites with most of them $\leq 0.9$.  We also investigated the scenarios of a  correlation of 0.3 between $T_0$ and $T_1$. The results are similar to the scenarios of a correlation of 0.5 (results not shown).


	\begin{threeparttable}[h!]%
		\centering
		\caption{\small Simulation results for estimators of the survival function of GMI, i.e. $S_{{\rm G}}(r)$, where $T_1$ and $T_0$ were generated from a Weibull frailty model with a correlation of 0.5 and ratio of medians equal to 1.3. \label{table:simu2}}%
		{\scriptsize \begin{tabular}{lllllllllllllllllllllllllllllll}
\hline
&&&& \multicolumn{4}{c}{Proposed} & \multicolumn{4}{c}{KM} & \multicolumn{4}{c}{lognormal} & \multicolumn{4}{c}{loglogistic} \\
$\sigma$ &  r  & C & n & Bias & SE & SEE & CP & Bias & SE & SEE & CP & Bias & SE & SEE & CP & Bias & SE & SEE & CP\\
  \cline{1-20} .3 & 1.3 &20\% &50 & .009 & .076 & .076 & .95 & .022 & .074 & .072 & .94 & .034 & .061 & .059 & .92 & .03 & .066 & .064 & .93\\
 &  & &70 & .01 & .063 & .064 & .96 & .025 & .06 & .061 & .95 & .034 & .05 & .05 & .91 & .031 & .053 & .054 & .92\\
 &  & &90 & .012 & .056 & .057 & .95 & .027 & .054 & .054 & .93 & .036 & .045 & .044 & .88 & .033 & .048 & .048 & .9\\
 &  &30\% &50 & .015 & .079 & .08 & .96 & .038 & .076 & .074 & .93 & .053 & .063 & .061 & .88 & .049 & .067 & .066 & .9\\
 &  & &70 & .018 & .068 & .067 & .95 & .041 & .063 & .063 & .92 & .053 & .052 & .052 & .84 & .05 & .055 & .056 & .87\\
 &  & &90 & .018 & .061 & .06 & .94 & .043 & .056 & .056 & .9 & .055 & .047 & .045 & .8 & .051 & .05 & .049 & .84\\
\cline{3-20} & 1.5 &20\% &50 & .013 & .078 & .076 & .95 & .033 & .075 & .073 & .94 & .052 & .063 & .061 & .88 & .039 & .067 & .066 & .92\\
 &  & &70 & .014 & .063 & .065 & .96 & .035 & .06 & .062 & .92 & .053 & .052 & .052 & .84 & .039 & .055 & .055 & .91\\
 &  & &90 & .014 & .057 & .057 & .95 & .037 & .055 & .055 & .91 & .055 & .047 & .046 & .8 & .041 & .049 & .049 & .88\\
 &  &30\% &50 & .024 & .083 & .081 & .94 & .055 & .079 & .076 & .9 & .074 & .066 & .065 & .81 & .061 & .071 & .069 & .87\\
 &  & &70 & .024 & .069 & .069 & .95 & .056 & .064 & .065 & .88 & .075 & .055 & .054 & .73 & .062 & .058 & .058 & .83\\
 &  & &90 & .024 & .064 & .061 & .93 & .058 & .058 & .057 & .84 & .077 & .05 & .048 & .66 & .063 & .053 & .051 & .78\\
\cline{3-20} & 1.7 &20\% &50 & .016 & .075 & .073 & .94 & .042 & .073 & .071 & .91 & .062 & .064 & .061 & .83 & .042 & .065 & .063 & .9\\
 &  & &70 & .014 & .062 & .062 & .95 & .041 & .06 & .06 & .91 & .062 & .053 & .052 & .78 & .042 & .053 & .054 & .89\\
 &  & &90 & .016 & .055 & .055 & .94 & .043 & .054 & .053 & .88 & .064 & .047 & .046 & .72 & .044 & .048 & .047 & .86\\
 &  &30\% &50 & .03 & .081 & .081 & .94 & .067 & .079 & .076 & .87 & .086 & .068 & .066 & .76 & .066 & .07 & .069 & .85\\
 &  & &70 & .026 & .068 & .069 & .94 & .065 & .065 & .065 & .84 & .087 & .056 & .055 & .67 & .067 & .058 & .058 & .81\\
 &  & &90 & .027 & .063 & .061 & .93 & .067 & .058 & .057 & .79 & .088 & .051 & .049 & .58 & .068 & .053 & .051 & .74\\
 \cline{1-20} .5 & 1.3 &20\% &50 & .009 & .075 & .076 & .96 & .022 & .073 & .072 & .95 & .033 & .061 & .059 & .92 & .03 & .065 & .064 & .94\\
 &  & &70 & .011 & .063 & .064 & .95 & .026 & .061 & .061 & .95 & .034 & .05 & .05 & .91 & .031 & .054 & .054 & .92\\
 &  & &90 & .011 & .057 & .057 & .95 & .025 & .055 & .054 & .93 & .035 & .044 & .044 & .89 & .032 & .047 & .048 & .91\\
 &  &30\% &50 & .019 & .081 & .079 & .96 & .04 & .077 & .074 & .93 & .053 & .063 & .061 & .88 & .049 & .067 & .066 & .9\\
 &  & &70 & .021 & .068 & .067 & .95 & .043 & .063 & .063 & .93 & .053 & .052 & .052 & .84 & .051 & .055 & .056 & .88\\
 &  & &90 & .02 & .059 & .06 & .95 & .042 & .056 & .056 & .9 & .054 & .045 & .045 & .81 & .051 & .048 & .049 & .84\\
\cline{3-20} & 1.5 &20\% &50 & .012 & .077 & .076 & .95 & .029 & .074 & .073 & .94 & .045 & .062 & .061 & .9 & .035 & .066 & .066 & .93\\
 &  & &70 & .013 & .063 & .065 & .96 & .031 & .061 & .062 & .94 & .046 & .052 & .051 & .87 & .036 & .055 & .055 & .91\\
 &  & &90 & .013 & .058 & .057 & .95 & .032 & .055 & .055 & .92 & .047 & .045 & .045 & .84 & .037 & .048 & .049 & .88\\
 &  &30\% &50 & .024 & .083 & .081 & .94 & .05 & .079 & .076 & .91 & .066 & .065 & .063 & .84 & .057 & .07 & .068 & .89\\
 &  & &70 & .025 & .068 & .069 & .95 & .052 & .064 & .064 & .9 & .067 & .054 & .053 & .78 & .058 & .057 & .058 & .85\\
 &  & &90 & .024 & .061 & .061 & .94 & .052 & .057 & .057 & .88 & .068 & .047 & .047 & .73 & .059 & .05 & .051 & .81\\
\cline{3-20} & 1.7 &20\% &50 & .014 & .078 & .076 & .95 & .035 & .075 & .073 & .93 & .053 & .063 & .061 & .87 & .038 & .066 & .065 & .93\\
 &  & &70 & .014 & .064 & .064 & .95 & .035 & .062 & .062 & .92 & .054 & .053 & .052 & .82 & .04 & .055 & .055 & .9\\
 &  & &90 & .014 & .058 & .057 & .94 & .037 & .055 & .055 & .91 & .056 & .046 & .046 & .78 & .041 & .049 & .049 & .87\\
 &  &30\% &50 & .028 & .084 & .082 & .94 & .057 & .08 & .077 & .89 & .076 & .067 & .065 & .79 & .062 & .071 & .069 & .87\\
 &  & &70 & .027 & .068 & .07 & .95 & .058 & .065 & .065 & .87 & .077 & .056 & .055 & .71 & .063 & .059 & .058 & .83\\
 &  & &90 & .027 & .062 & .062 & .93 & .059 & .058 & .057 & .84 & .078 & .048 & .048 & .66 & .064 & .051 & .051 & .79\\
\cline{1-20}
\end{tabular}}
		\begin{tablenotes}
			\item{\small Note: C, censoring rate; Bias, the sampling bias; SE, the sampling standard error; SEE, the sampling mean of the standard error estimator;
CP, the coverage probability of the 95\% confidence interval. Each entry was based on 2000 repliates. SEEs for the proposed estimator and parametric estimators were calculated based on 5000 bootstrap samples.}
		\end{tablenotes}
	\end{threeparttable}

\section{Example}
We considered data from a phase II clinical trial on advanced colorectal cancer patients \cite{bonetti2001use, kovalchik2011statistical, wu2019phase}. This trial used GMI, i.e. the ratio of TTP during first- and second-line therapy, as the primary endpoint to measure the activity of the combination of oxaliplatin with 5-fluorouracil compared to 5-fluorouracil alone. The first-line therapy was 5-fluorouracil and the second-line of therapy was the combination of oxaliplatin with 5-fluorouracil. Patients were switched to second-line therapy after they progressed following first-line therapy. The TTP under first-line therapy was observed on the enrolled 34 patients, and the TTP under second-line therapy was oberved on 27 (79\%) patients and censored on the rest 7 (21\%) patients. A primary question of the study is to estimate the distribution of GMI. Particularly, the investigators are interested in estimating  probability that GMI greater than a threshold of 1.3\cite{bonetti2001use}. 

\begin{figure}
\centering
\includegraphics{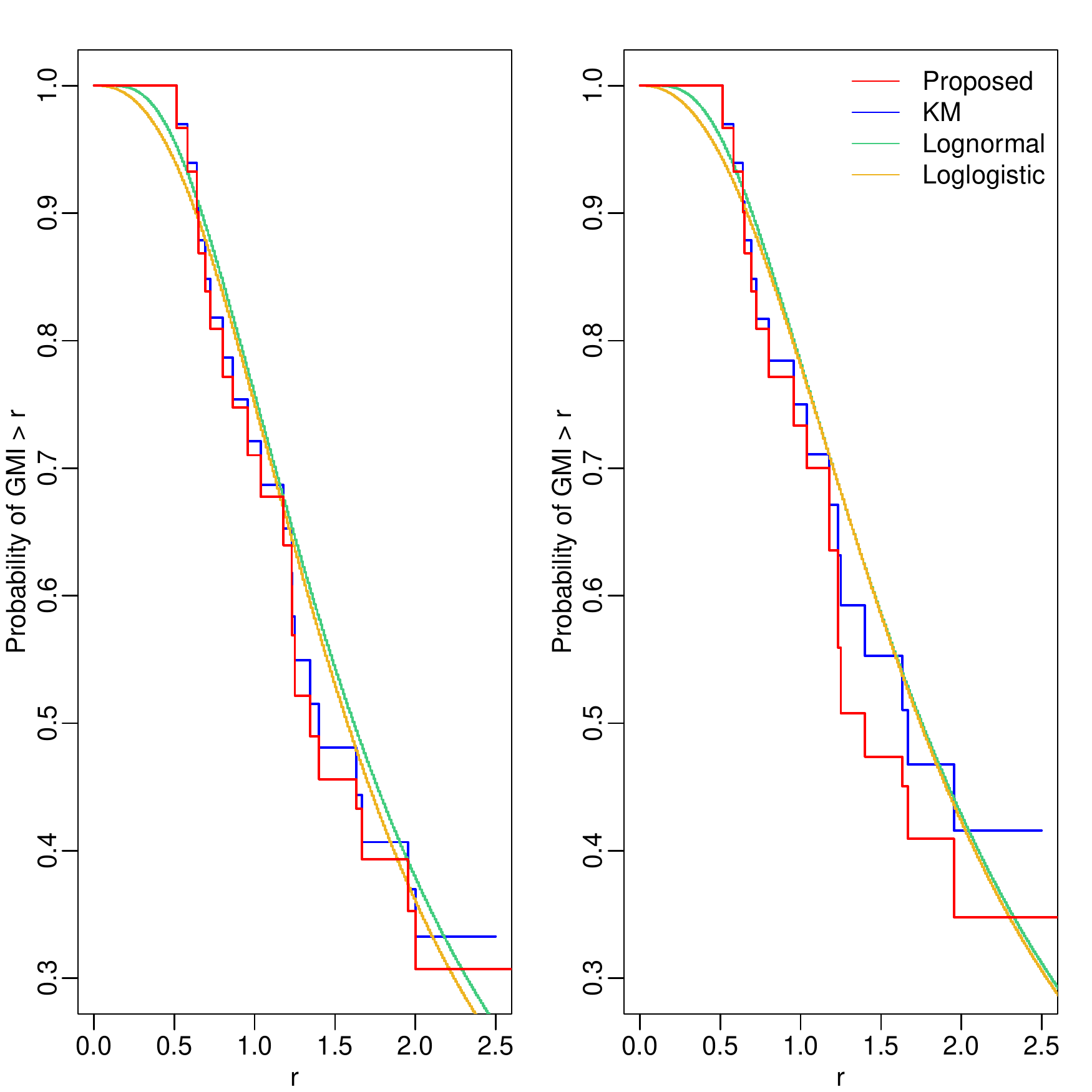}
\caption{Estimated survival functions of GMI based on the original data (left panel) and data with increased censoring that additionally restricted the follow-up time of each patient to 9 months (right panel).}
\label{fig:real}
\end{figure}

We applied the proposed estimator, $\hat{S}_{{\rm G}}(\cdot)$, to estimate the survival function of GMI and compared to the KM estimator and two parametric estimators based on the lognormal  and loglogistic distributions, respectively. The left panel of Figure \ref{fig:real} displays the estimates of the survival function of GMI based on those methods. The estimated curves based on the KM method and parametric methods, especially the latter, were considerably higher than that based on our proposed method. Therefore, the KM  and parametric methods overestimated the survival function of GMI. To examine the impact of censoring rate on the performance of different methods, we additionally restricted the follow-up time of each patient to 9 months, which yielded an increased censoring rate of 32\%. As shown in the right panel of Figure \ref{fig:real}, the difference between the estimated curves based on our method and  KM and parametric methods became larger. Table \ref{table:real} reports the estimated probabilities of GMI greater than several selected threshold values including 1.3, 1.5 and 1.7. The relative differences between other methods and our method ranged from 3\% to 19\% for the original data. For the data with censoring rate increased to 32\%, the relative differences increased to 14\% to 30\%. Compared to our method, the KM method significantly overestimated the survival function of GMI at threshold values of 1.3 and 1.5. The paramteric methods also significantly overestimated the survival function of GMI at threshold value of 1.3.  

\begin{center}
	\begin{threeparttable}[h!]%
		\centering
		\caption{\small Estimated probability of GMI $>$ 1.3, 1.5 or 1.7. \label{table:real}}%
	{\footnotesize	\begin{tabular*}{500pt}{@{\extracolsep\fill}lccccccccc@{\extracolsep\fill}}
\toprule
& & \multicolumn{8}{c}{Original data}\\
& & \multicolumn{2}{c}{1.3}& & \multicolumn{2}{c}{1.5}& & \multicolumn{2}{c}{1.7}\\
 \cline{3-4} \cline{6-7} \cline{9-10}
Method & & Est (\%Dif) & p-value   & & Est (\%Dif) & p-value   & & Est (\%Dif) & p-value  \\
\hline Proposed & & 0.52 (~---~) & ~--- & & 0.46 (~---~) & ~--- & & 0.39 (~---~) & ~--- \\
\hline KM & & 0.55 (~5.3) & 0.20 & & 0.48 (~5.4) & 0.25 & & 0.41 (~3.4) & 0.56 \\
\hline lognormal & & 0.62 (19.3) & 0.07 & & 0.54 (18.8) & 0.12 & & 0.47 (19.4) & 0.15 \\
\hline loglogistic & & 0.61 (17.4) & 0.08 & & 0.53 (15.9) & 0.14 & & 0.45 (15.3) & 0.22 \\
\hline & & \multicolumn{8}{c}{Data with increased censoring}\\
& & \multicolumn{2}{c}{1.3}& & \multicolumn{2}{c}{1.5}& & \multicolumn{2}{c}{1.7}\\
 \cline{3-4} \cline{6-7} \cline{9-10}
Method & & Est (\%Dif) & p-value   & & Est (\%Dif) & p-value   & & Est (\%Dif) & p-value  \\
\hline Proposed & & 0.51 (~---~) & ~--- & & 0.47 (~---~) & ~--- & & 0.41 (~---~) & ~--- \\
\hline KM & & 0.59 (16.7) & 0.02 & & 0.55 (16.7) & 0.03 & & 0.47 (14.2) & 0.11 \\
\hline lognormal & & 0.66 (30.0) & 0.02 & & 0.58 (23.4) & 0.09 & & 0.52 (26.1) & 0.09 \\
\hline loglogistic & & 0.66 (30.0) & 0.02 & & 0.58 (23.2) & 0.07 & & 0.51 (25.4) & 0.07 \\
\bottomrule
		\end{tabular*}}
		\begin{tablenotes}
			\item{\small Note: Est is the estimated probability; \%Dif is the difference in estimated probabilities between a traditional estimator and the proposed estimator relative to the proposed estimator; and p-value is based on a Wald test calculated as the difference between a traditional estimator and the proposed estimator divided by the standard error of the difference obtained using bootstrap.}
		\end{tablenotes}
	\end{threeparttable}
\end{center}

\section{Discussion}
To our knowledge, we are the first to identify the dependent censoring issue of GMI. Traditional methods for right-censored data are not appropriate for GMI data because they require the independent censoring assumption. Our simulation studies further demonstrate that traditional methods for estimating the survival function of GMI, including the KM method and parametric lognormal and loglogistic methods, can lead to severely biased results in certain situations. Therefore, it is critical to take into account dependent censoring when developing statistical methods for GMI data. 

We have developed consistent nonparametric estimators for the survival function of GMI. A key step in the development of the esitmators is to estimate the conditional survival function of GMI. We utilized the kernel method to estimate the conditional survival function so that the estimators for the survival function of GMI are purely nonparametric. A common question for using the kernel method is how to choose the bandwidth. We derived the conditions that the bandwidth needs to satisfy so that the estimators are consistent and weakly convergent. We considered three bandwidths that satisfy those conditions. The simulation results based on the bandwidth of $\hat{\sigma}_0 n^{-2/5}$ were presented in the paper. The results based on bandwiths of $\hat{\sigma}_0 n^{-1/3}$ and $\hat{\sigma}_0 n^{-1/2}$ were similar (data not shown). Thus, our estimator is robust to the choice of bandwidth. 

We have focused on the estimation of the survival function of GMI, i.e. the probability of GMI greater than a given value. It is also of interest to  assess the treatment efficacy by hypothesis testing. In practice, one approach is to choose a threshold and test whether the probability of GMI greater than that threshold is greater than a prespecified proportion \cite{von1998there}. Frequently used thresholds are 1.3 and 1.5 \cite{von1998there, radovich2016clinical, rodon2019genomic}, where GMI greater than those values are considered as clinically meaningful improvement of the new treatment, although those threshold values are arbitrary \cite{von1998there}. To test such a hypothesis,  Wald tests can be constructed based on our estimators. Because our simulation studies show that our estimators are unbiased and confidence intervals have proper coverage probabilities, the Wald tests based on our estimators are expected to have controlled type I errors. In contrast, the Wald tests based on KM and parametric lognormal and loglogistic estimators will have inflated type I errors because those estimators have low coverage probabilities in many situations. 



There is also a great interest in comparing the survival functions of GMI between two groups. For example, in a phase II trial evaluating the clinical benefit of genomically guided therapy\cite{radovich2016clinical}, researchers were interested in comparing the survival functions of GMI  at threshold values such as 1.3 and 1.5 and regardless of threholds between patients receiving genomically guided therapy and patients receiving non-geonmically guided therapy. For such a problem, Wald tests can be constructed based on our estimators to compare the survival functions of GMI  at a given threshold between the two groups. Our estimators also allow virtually comparing survival functions of GMI at all possible threshold values. A future work of interest is to develop an overall statistical test to compare two survival functions of GMI across all thresholds.


\section*{Acknowledgments}
This research was partially supported by  the Biostatistics and Bioinformatics Shared Resource Facility of the
University of Kentucky Markey Cancer Center (P30CA177558). 

\section*{Appendix: Asymptotic properties of $\hat{S}_{{\rm G}}(r) $ and $\tilde{S}_{{\rm G}}(r) $}
Using the modern empirical process theory, we first prove the consistency and weak convergence of $\tilde{S}_{{\rm G}}(r)$, and then obtain these asymptotic properties of the simpler estimator $\hat{S}_{{\rm G}}(r)$ following a similar procedure. Let $P_n$ and $P$ denote the empirical measure and the distribution under the true model, respectively. For a measurable function $f$ and measure $Q$, the integral $\int f dQ$ is abbreviated as $Qf$. 

It is useful to adopt the counting process notation. Let $N(s) = I(Y_{1} / T_{0} \leq s, \Delta_{1}=1)$ be the observed counting process for $T_{1} / T_{0}$, and $A(s) = I(Y_{1} / T_{0} \geq s)$ be the corresponding at-risk process. Let $W=(\log T_0, Z, V)$, $\sqrt{n}\{\tilde{S}_{{\rm G}}(r) -{S}_{{\rm G}}(r)\}$ can be written as $\sqrt{n}\{P_n\tilde{S}_{{\rm G}|W}(r; W)-P S_{{\rm G}|W}(r; W)\}$, where $S_{{\rm G}|W}(r;w) = {\rm Pr}(T_1/T_0>r | W=w)$ and
\begin{equation*}
\tilde{S}_{{\rm G}|W}(r; W) = \Pi_{0<s\leq r} \left[1-\frac{\sum_{j=1}^n \tilde{K}\left\{\left(W_j - w\right)/{a_n}\right\} dN_j(s)}{\sum_{j=1}^n \tilde{K}\left\{\left(W_j - w\right)/{a_n}\right\} A_j(s)}\right].
\end{equation*}
Here, $N_j(s)$ and $A_j(s)$ are the observed counting process and at-risk process for the $j$th individual, respectively, and $\tilde{K}(u) = K(\left\|(u_1)\right\|)I(u_2=0)$, where $u_1$ is the first $q+1$ elements of $u$ and $u_2$ are the rest elements.

We further express $\sqrt{n}P_n\tilde{S}_{{\rm G}|W}(r; W)-P S_{{\rm G}|W}(r; W)$ as
\begin{equation} \label{eq:S}
 \sqrt{n}(P_n-P)\{S_{{\rm G}|W}(r; W) - S_{\rm G}(r)\}+\sqrt{n}P\{\tilde{S}_{{\rm G}|W}(r; W)-S_{{\rm G}|W}(r; W)\} \\
+ \sqrt{n}(P_n-P)\{\tilde{S}_{{\rm G}|W}(r; W)-S_{{\rm G}|W}(r; W)\}.
\end{equation}
Based on the result for the second term in equation (A6) of \cite{chen2012predictive} with $c_v=\infty$,  the second term in equation (\ref{eq:S}) is asymptotically equivalent to 
\begin{equation*}
\sqrt{n}(P_n-P)\left[-S_{{\rm G}|W}(r; W) \int_0^r\frac{dN(s)+A(s)d\log S_{{\rm G}|W}(s; W)}{E\{A(s)|W\}}\right],
\end{equation*}
under the conditions that 1)
there exists some integer $l$ such that $\int u^{j} K(u)dy=0$ ($j=1, \ldots, l-1$),
$\int u^{l} K(u)du\neq 0$ and $2l-q-1>0$, and 2)
$na_n^{q+1}\rightarrow \infty$ and $na_n^{2l}\rightarrow 0$ as $n\rightarrow\infty$. When $q\leq 2$, we can use a symmetric smooth probability density function as the kernel function $K(\cdot)$.

Similarly, we can verify that $ P\{\tilde{S}_{{\rm G}|W}(r; W)-S_{{\rm G}|W}(r; W)\}^2\longrightarrow_{p}0$ uniformly for $t\in[0,\infty] $ and  that $\tilde{S}_{{\rm G}|W}(r; W)$ and $S_{{\rm G}|W}(r; W)$ belong to a $P$- Donsker class. It then follows that the third term of equation (\ref{eq:S}) converges uniformly to zero in probability by Lemma 19.24 of \cite{van1998asymptotic}.

Combining the aforementioned results, we obtain that $\sqrt{n}\{\tilde{S}_{{\rm G}}(r) -{S}_{{\rm G}}(r)\}$ is asymptotically equivalent to the process
\begin{align*}
& \sqrt{n}(P_n-P)\left[S_{{\rm G}|W}(r; W) - S_{\rm G}(r)-S_{{\rm G}|W}(r; W) \int_0^r\frac{dN(s)+A(s)d\log S_{{\rm G}|W}(s; W)}{E\{A(s)|W\}}\right].\\
=& n^{-1/2} \sum_{i=1}^n \left[- S_{\rm G}(r) + S_{{\rm G}| T_0, Z, V}(r; T_{0i}, Z_{i}, V_i)\left\{1-\frac{\Delta_{1i}I(Y_{1i}/T_{0i} \leq r)}{H_c(Y_{1i}/T_{0i}; T_{0i}, Z_i, V_i)} \right.\right. \\
&\hspace{7cm}- \left.\left.\int_0^r\frac{I(Y_{1i}/T_{0i} \leq s)d\log S_{{\rm G}| T_0, Z, V}(s; T_{0i}, Z_{i}, V_i)}{H_c(s; T_{0i}, Z_i, V_i)}\right\}  \right],
\end{align*}
where $H_c(s; t_0, z, v) = {\rm Pr}(Y_{1}/T_{0}\geq s|T_0=t_0, Z=z, V=v)$. Therefore, $\tilde{S}_{{\rm G}}(r)$ is consistent and $\sqrt{n}\{\tilde{S}_{{\rm G}}(r) -{S}_{{\rm G}}(r)\}$ is weakly convergent to a zero-mean Gaussian process.

Following a similar procedure as above with $W = \log T_0$, we obtain that $\sqrt{n}\{\hat{S}_{{\rm G}}(r) -{S}_{{\rm G}}(r)\}$ is asymptotically equivalent to the process
\begin{align*}
 n^{-1/2} \sum_{i=1}^n \left[S_{{\rm G}| T_0}(r; T_{0i})\left\{1-\frac{\Delta_{1i}I(Y_{1i}/T_{0i} \leq r)}{H(Y_{1i}/T_{0i}; T_{0i})}  - \int_0^r\frac{I(Y_{1i}/T_{0i} \leq s)d\log S_{{\rm G}| T_0}(s; T_{0i})}{H(s; T_{0i})}\right\} - S_{\rm G}(r) \right],
\end{align*}
under the conditions that $K(\cdot)$ is a symmetric smooth probability density function and $na_n\rightarrow \infty$ and $na_n^4\rightarrow 0$. Therefore, $\hat{S}_{{\rm G}}(r)$ is consistent and $\sqrt{n}\{\hat{S}_{{\rm G}}(r) -{S}_{{\rm G}}(r)\}$ is weakly convergent to a zero-mean Gaussian process.

\nocite{*}
\bibliography{pfs}
\bibliographystyle{WileyNJD-AMA}

\end{document}